\documentstyle[preprint,aps,pre]{revtex}

\begin{document}

\draft

\preprint{cond-mat/9402010}

\title{Properties and Origins of Protein Secondary Structure}

\author{Nicholas D. Socci$^{(1)}$,
	William S. Bialek$^{(2)}$,
	and Jos\'e Nelson Onuchic$^{(1)}$}

\address{$^{(1)}$Department of Physics, University of California at San Diego,
La Jolla, California 92093\\
$^{(2)}$NEC Research Institute, Princeton, New Jersey 08540}

\date{\today}

\maketitle

\begin{abstract}

Proteins contain a large fraction of regular, repeating conformations,
called secondary structure. A simple, generic definition of secondary
structure is presented which consists of measuring local correlations
along the protein chain. Using this definition and a simple model for
proteins, the forces driving the formation of secondary structure are
explored. The relative role of energy and entropy are examined. Recent
work has indicated that compaction is sufficient to create secondary
structure. We test this hypothesis, using simple non-lattice protein
models.

\end{abstract}


\narrowtext

Recently, there has been a great deal of interest in the study of
proteins from a physical
perspective~\cite{Chan93,Bascle93a,Iori92,Leopold92,Shakhnovich91,Sasai90}.
Most of these works have focused on the folding problem; {\em i.e.\/},
how does the sequence of amino acids encode the three-dimensional
structure of the protein. Although progress has been made in this
area, there is still a long way to go before there is a complete
understanding of how proteins fold. However, proteins have many other
interesting properties. While each protein has a specific structure
determined by its sequence, all proteins share several common
structural features. They are highly compact, with very little free
internal space. More striking is the high degree of order found, which
consists of regular periodic arrangements of the main chain into one
of a few universal patterns (called {\em secondary structure}).
Roughly 50\% of the structure of all proteins is in some form of
secondary structure~\cite{Kabsch83}. In this paper we define in a
simple, generic way precisely what secondary structure is. This
definition will be valid not only for proteins but for simpler
polymers and simple protein like models. We then use it to investigate
what forces are responsible for the formation of secondary structure.
Although this is not directly related to the folding problem, a
thorough understanding of what factors are responsible for secondary
structure may aid in the study of the folding problem.

There has been a great deal of past work attempting to understand the
origins of secondary structure. At first it was believed that {\em
local} interactions (local hydrogen-bonds or dihedral angle potentials
for example) were responsible. Here, the term local means close with
respect to the separation along the polymer chain. For example, a
hydrogen bond between monomer $i$ and $i+4$ would be a local
interaction, as would an angle potential. Several recent studies
indicate that local forces may not be the dominant effect, rather
compaction of the chain may be the important factor. By examining
exhaustive enumerations of short chains on a lattice, Chan and
Dill~\cite{Chan91a,Chan90b,Chan89a} found that as the compactness of
the chains increased so did the percentage of secondary structure
present.  They also found that the maximally compact chains had
roughly the same amount of secondary structure as real proteins and
the proportions of helices to sheets was also approximately the same.
Subsequently, Gregoret and Cohen~\cite{Gregoret91} studied non-lattice
models. Their results also suggest that compactness does influence the
amount of secondary structure, but they indicate that the effect is
most pronounced at densities 30\% greater than that of real proteins.
In both of these studies however, local interactions were present.
For example, a lattice has a specific set of allowed bond angles,
which provides an effective bond angle potential. In the non-lattice
work, compact chains were generated using a biased random walk in
which the bond angles were chosen not from a uniform distribution but
from the distribution observed in real-proteins. This also provides an
effective angle potential.  Therefore, it is not clear from these
works whether compaction is sufficient to generate secondary
structure. We wish to determine whether compaction, without local
interactions, is sufficient.

There are two distinct questions to keep in mind: (1) why do proteins
(or other polymers) form regular structures and (2) why do proteins
form particular types of secondary structure. Question one is
equivalent to asking, why do proteins form helices and sheets. The
second question asks, why are these helices $\alpha$-helices and the
sheets $\beta$-sheets. The answer to the second question certainly
involves local interactions. It is the specific hydrogen bonding
patterns in proteins which favor the formation of $\alpha$-helices. In
other polymers, different local interactions would favor other forms.
For example, the structures of 179 polymers have been solved and 79
are found to be in one of 22 different types of
helices~\cite{Chan90b,Tadokoro79}. In each polymer the specific types
of local interactions determine the preferred type of secondary
structure. In this work we are interested in studying the first
question: what forces are responsible for formation of regular
structures. Specifically we will test the previous suggestions that
compaction of the chain is the key driving force. To do so we will be
using models without any local interactions. However, without local
interactions there is no way of knowing before hand what types of
secondary structure will be formed. Most definition of secondary
structure are specific to a given type of structure ({\em i.e.}
$\alpha$-helices), consequently one needs to know {\em a priori} what
types of secondary structures will occur in order to detect their
presence. To overcome this problem we developed a generic method of
determining whether secondary structure is present without the need to
know {\em a priori} what its specific form is.

A simple way of defining secondary structure is to realize that it
consists of repeating patterns.  Consequently the polymer chain should
be correlated with itself along the chain. The correlation length
should be related to the average size of secondary structures. To
detect secondary structure we measure the correlations between
different points along the protein chain.  Specifically, let
$\theta_j$ represent the value of the dihedral angle associated with
the $j^{\text{th}}$ $\alpha$-carbon (see figure~\ref{fig:dihedral}).
We then calculate:
\begin{equation}
{\cal C}_\theta(\Delta)
	= \left\langle e^{i(\theta_j -
\theta_{j+\Delta})}\right\rangle_{\text{C}}.
\label{eq:corr}
\end{equation}
The average is over $j$; that is, over all pairs of angles separated
by a distance $\Delta$ along the chain. The subscript $\text{C}$
indicates that the mean,
$\left|\left\langle e^{i\theta_j}\right\rangle\right|$,
has been subtracted from $\left\langle e^{i(\theta_j -
\theta_{j+\Delta})}\right\rangle$.
If secondary structure is present then ${\cal C}_\theta(\Delta)$ will
be non-zero for $\Delta\lesssim l_{\text{avg}}$ where $l_{\text{avg}}$
is related to the average length of secondary structure. Note, this
definition makes no reference to any particular type of secondary
structure; therefore, any form of regular structure will be detected.
For example, if helices are present there will be a non-zero
correlation length no matter what period the helices have.
Equation~\ref{eq:corr} also has the advantage that it can be
calculated analytically in a simple model.

To test our definition we examined the crystal structures from 112
proteins which have been recorded in the Protein Data
Bank (PDB)~\cite{Bernstein77}. The correlation function was calculated
for each protein and normalized so ${\cal C}_\theta(0)=1$.
Then an average correlation function was computed for all
proteins. Examining this correlation function (shown in
figure~\ref{fig:corr}) we see that protein chains are positively
correlated up to separations of approximately nine monomers. This is
comparable to the average length of secondary structure (roughly ten
monomers) measured by others~\cite{Kabsch83}. At distances greater
than nine monomers the chains become negatively correlated. This
negative correlation may be partly due to {\em supersecondary}
structure, which consists of combinations of secondary structural
elements. For example, $\beta$-sheets are usually followed by reverse
turns. There is also the $\beta\xi\beta$-unit where two parallel
$\beta$-sheets are separated by some piece $\xi$ which can be a random
coil, an $\alpha$-helix or another sheet~\cite{Schulz79}. Eventually
the correlations fall off to zero (at around $\Delta=16$).

We now examine what forces drive the formation of secondary structure,
specifically the question of whether the loss of entropy due to
compaction is sufficient. To do this we need a model without any local
interactions. Lattice models are not acceptable since the restricted
degrees of freedom imply local bond angle potentials. An off-lattice
model was used instead. As in lattice and other simple models we
neglect the internal degrees of freedom of the amino acids and
represent each as a single point in space. Monomers that are connected
along the chain are constrained to be separated by a fixed distance.
The next step is to fold the chains into compact conformations. The
following procedure was used. Take a potential energy function whose
minima are compact conformations. Then minimize this potential energy
to fold the chain. Because the model we are using is a {\em
homopolymer} there are many compact local minima (the number grows
exponentially with chain length~\cite{Chan89a}). We will generate an
ensemble of compact conformations, using chains of several different
lengths. One can think of this ensemble of different compact
structures as representing the collection of native structures of many
different sequences of amino acids. We will calculate the average
correlation function (eq.~\ref{eq:corr}) of the ensemble of compact
conformations we generate and look for long range correlations which
will indicate the presence of secondary structure. It is important to
note that the previous works showing the connection between compaction
and secondary structure~\cite{Chan91a,Chan90b,Chan89a,Gregoret91} also
used a homopolymer model and many homopolymers show secondary
structure in their compact states~\cite{Tadokoro79}. Therefore, it
does not appear necessary to have a heteropolymer and a unique ground
state to get secondary structure.

There are several different potentials that have compact minima. The
dominant force for the folding of proteins is the {\em hydrophobic
effect}~\cite{Dill90c}. This is primarily a bulk, entropic effect
caused by interactions of the polymer with the surrounding water.  The
protein collapses to create a hydrophobic core with polar groups on
the surface. One could simulate a polymer in a solution of water,
however, this is much more complex than necessary. Instead of doing a
full water-polymer simulation we simply choose an effective potential
which will also cause the polymer to collapse. The particular one used
in this work was:
\begin{eqnarray}
V\left(\left\{\vec r_i\right\}\right) &=&
	\sum_{i=1}^{N-1}\frac{1}{2}k_{\text{c}}
             \left(\left|\vec r_i-\vec r_{i+1}\right|
		-l_{\text{c}}\right)^2\nonumber\\
      &+&\epsilon\left\{
	 \sum_{i<j}^N\left(
	\frac{\sigma_{\text{ev}}}{r_{ij}}\right)^{12}
      -\frac{1}{N}\sum_{i=1}^N\left|\vec r_i
	- \vec r_{\text{com}}\right|^2\right\},
\label{eq:potential}
\end{eqnarray}
%
where $r_{ij}=\left|\vec r_i - \vec r_j\right|$, $\vec r_i$ is the
position of the $i^{\text{th}}$ monomer and $\vec
r_{\text{com}}=\frac{1}{N}\sum\vec r_i$ is the position of the center
of mass. The first term represents the covalent forces that bind the
monomers along the chain. The constants $k_{\text{c}}$ and
$l_{\text{c}}$ are both set equal to one, determining the energy and
length units. The middle term (which is the repulsive part of a
Lennard-Jones potential) is the excluded volume term which prevents
the chain from compacting to a single point.  The last term is the
radius of gyration of the chain. This term provides the compacting
force. The two constants, $\epsilon$ and $\sigma_{\text{ev}}$, are
determined by examining real proteins.  The difference in energy
scales between covalent and non-covalent forces determines $\epsilon$.
In proteins the typical non-covalent interaction is roughly
one-hundredth the energy of a covalent bond, so $\epsilon$ is set
equal to 0.01~\cite{Note02}. The compactness of the chains will be
controlled by the value $\sigma_{\text{ev}}$. To determine the value
of $\sigma_{\text{ev}}$ and measure compactness we looked at two
features of real protein structure: the pair-correlation function
(also called the radial distribution function) and the radius of
gyration. First, the pair-correlation function was measured for both
real proteins and our chains. This function gives the probability that
two $\alpha$-carbons are separated by a given distance, indicating how
closely the $\alpha$-carbons are packed together. We adjusted
$\sigma_{\text{ev}}$ until the position of the nearest neighbor peak
for our chains closely matched the one for real
proteins~\cite{Socci92}. Next, we measured the radius of gyration as a
function of chain length for real proteins. Our chains had a slightly
smaller radii of gyration as proteins the same length (see
figure~\ref{fig:radgyr}).  This is not surprising since the potential
we used will generate nearly spherical shapes while proteins are
ellipsoidal with varying eccentricities. An ellipsoid will have a
larger radius of gyration than a sphere of equal volume.

The chains were compacted by minimizing this potential energy
(equation~\ref{eq:potential}). The algorithm used was a
conjugate-gradient decent minimizer~\cite{Press86}. At each iteration
in this algorithm the energy is decreased, so it is somewhat analogous
to a zero temperature Monte-Carlo simulations, in that only energy
reducing steps are accepted. There is the possibility that for some
potentials this type of algorithm will be trapped in local non-compact
minima. However for the potential used here this was not a problem.
All minima that we generated were observed to be compact; {\em i.e.},
their radius of gyration was roughly the same as those of proteins the
same length (see figure~\ref{fig:radgyr}). Starting from a random
initial condition (which was a taken to be a self-avoiding random
walk) 200 chains, ranging in length from 50 to 450
monomers~\cite{Note01}, were folded. The average dihedral angle
correlation function was then calculated for these chains to determine
if any secondary structure was present.  Figure~\ref{fig:rgtorcor}
shows the average for the compacted chains with the correlation
function for real proteins superimposed. The compacted chains show no
long range correlations.  The plot falls almost immediately to zero,
with a slight negative correlation at separations of roughly two
monomers. This lack of any correlations indicates the absence of any
secondary structure.

The potential (equation~\ref{eq:potential}) was chosen to have no
local interactions other than the one term which bonds a monomer to
its two neighbors along the chain. Again, local here means local
(close) as measured along the chain, not through space. The excluded
volume term is through space local, but in a folded structure any two
monomers can interact via the excluded volume term regardless of there
separation along the chain. In particular, there is no angle term in
the potential (either implicit or explicit). The previous works which
did find secondary structure with increasing compactness did have
implicit angle potentials. It appears that compacting the chain is not
enough to generate secondary structure. It is possible that the
particular form of the compacting potential we used destroys secondary
structure or was biased in favor of compact conformation without
secondary structure.

To test this we tried a different compacting potential, the
Lennard-Jones 6--12 potential. We replaced the radius of gyration term
in eq.~\ref{eq:potential} by a $r^{-6}$ term to give:
\begin{eqnarray}
V\left(\left\{\vec r_i\right\}\right) &=&
	\sum_{i=1}^{N-1}\frac{1}{2}k_{\text{c}}
             \left(\left|\vec r_i-\vec r_{i+1}\right|
		-l_{\text{c}}\right)^2\nonumber\\
      &+&\epsilon\left\{
	 \sum_{i<j}^N\left(
	\frac{\sigma_{\text{ev}}}{r_{ij}}\right)^{12}
      - \left(\frac{\sigma_{\text{ev}}}{r_{ij}}\right)^{6}\right\}.
\end{eqnarray}
By itself the 6--12
potential is too short ranged to compact an extended chain so we did a
two stage minimization. At the first we added an additional $1/r$
piece which is long ranged and will collapse an extended chain. Once
the chain was semi-compact, we finish the minimization without the
$1/r$ term. We generated an ensemble of compact chains and measured
the average correlation function (see figures~\ref{fig:radgyr}
and~\ref{fig:torcorr}). Again there where no long range correlations
hence no secondary structure.

To explore the forces responsible for the formation of secondary
structure in proteins we have defined a simple, generic method of
measuring secondary structure in polymers. This method consists of
calculating the angle correlation function along the chain and looking
for long range correlations. If secondary structure is present there
will be long range correlations with a length comparable to average
size of the secondary structure. This method does not depend on the
precise details of what type of structure is present and can be used
when these details are not known. Real proteins whose structures have
been solved were examined and long range correlations were found. This
technique was then used to examined whether compaction leads to the
formation of secondary structure. Simple models with no local
interactions were used and two different compacting potentials were
examined. There were no long range correlations indicating the absence
of secondary structure was present. These results indicate that
compaction by itself is not sufficient to generate secondary
structure. In the previous studies demonstrating a connection between
secondary structure and compaction there was always some form of local
interactions present.  It appears, however, that local interactions are
not sufficient since compactness was also necessary to get structure.
In proteins the formation of secondary structure appears to result
from the combination of both the entropic effect of compaction and
local energetic effects. The loss of entropy from compaction is not
enough to force the chain into regular conformations. Using our
definition of secondary structure further studies can be carried out
to determine the relative importance of the these two factors.

We acknowledge helpful discussions with S.~Skourtis, A.~Libchaber,
A.~Schweitzer and S.~Favarolo. J.~N.~O. is a Beckman Young
Investigator. This work was funded by the Arnold and Mabel Beckman
Foundation and the National Science Foundation (Grant No.  MCB-9018768
and a previous graduate fellowship to N.~D.~S.). J.~N.~O. is in
residence at the Instituto de F\'{\i}sica e Qu\'{\i}mica de S\~ao
Carlos, Universidade de S\~ao Paulo, S\~ao Carlos, SP, Brazil during
part of the summers.


\widetext

\begin{figure}[htb]
\caption{The dihedral (also called torsion) angle, $\Theta_i$,
associated with the $i^{\rm th}$ monomer. The inset shows the view
along the bond from monomer $i-1$ to $i$. The angle shown is defined
as positive by our sign convention.}
\label{fig:dihedral}
\end{figure}
\begin{figure}[htb]
\caption{Real part of the dihedral angle correlation function averaged
over 112 proteins from the protein data bank. The distance, $\Delta$,
is the number of monomers along the chain. ${\cal C}_\theta(0)$ has
been normalized to one.}
\label{fig:corr}
\end{figure}
\begin{figure}
\caption{The radius of gyration versus chain length (plotted on a
log-log scale) for real proteins (small circles), chains compacted
using the radius of gyration potential (diamonds), and the
Lennard-Jones potential (stars).  The radius of gyration for the three
systems is very similar indicating that they all have the same level
of compactness.  }
\label{fig:radgyr}
\end{figure}
\begin{figure}
\caption{The two solid lines show the correlation functions for the radius of
gyration potential (circles) and Lennard-Jones potential (squares).
The dotted line is the real protein correlations (from
figure~\protect\ref{fig:corr}) for comparison.
}
\label{fig:rgtorcor}\label{fig:torcorr}
\end{figure}

\end{document}